\documentclass[prl,twocolumn,showpacs,preprintnumbers,amsmath,amssymb,floats]{revtex4}
\usepackage{graphicx}
\usepackage{dcolumn}
\usepackage{bm}

\usepackage{amsmath}
\usepackage{color}
\usepackage{soul}

\def\be{\begin{equation}}
\def\ee{\end{equation}}
\def\bea{\begin{eqnarray}}
\def\eea{\end{eqnarray}}

\begin{document}
\title{Magneto-oscillations in Underdoped Cuprates}
\author{C. M. Varma}
\affiliation{Department of Physics and Astronomy, University of
California, Riverside, California 92521}
\begin{abstract}
The conventional interpretation of the recent magneto-oscillation experiments in underdoped Cuprates, requires that there be a state of altered translational symmetry in the pseudogap state which is not supported by structural and Angle Resolved Photoemission Spectroscopy (ARPES) experiments. I show here that the observed oscillations  may be reconciled with the conclusion arrived in ARPES experiments that the fermi-surface, suitably defined, has the shape of four arcs which shrink to four points as $T \to 0$.  Experiments, including infrared absorption in a magnetic field, are suggested to distinguish between such a state from that obtained by the conventional interpretation of the magneto-oscillations.  
\end{abstract}
\maketitle

A series of remarkable experiments \cite{magneto-oscExpts.}, \cite{jaudet} on very high quality samples have recently reported observation of magneto-oscillations (M-O) in the "normal" state of under-doped cuprates at high enough magnetic fields $H$ to suppress superconductivity. The experiments have been interpreted in terms of the conventional Onsager, Lifshitz-Kosevich theory for normal metals to suggest a Fermi-surface cross-section of only a few per-cent of what is expected of the  Fermi-surface of optimally or overdoped cuprates. In the theory of usual metals, such a small Fermi-surface would require a change of translational symmetry from overdoped to underdoped cuprates such that the partially filled band has a much reduced number of carriers. 

The conventional picture appears not in accord with what is inferred through ARPES or indeed the thermodynamic experiments such as specific heat and magnetic susceptibility in the normal states.  ARPES shows the unusual phenomena of "Fermi-Arcs" \cite{fermiarcs}, below a temperature which (within some experimental uncertainly) coincides with $T^*(x)$, the temperature below which the transport and thermodynamic property of cuprates show characteristic changes. To the best resolution of the ARPES experiments, (peaks in the spectral-function of) one-particle states are found at the chemical potential only over an arc of the {\it expected Fermi-surface} with the angle 
$\Phi$ of the arc, measured from the $(\pi,\pi)$ direction, decreasing as temperature is decreased. Outside the arc there is a gap at the chemical potential. Superconductivity intervenes so that the Fermi-surface, which it is useful to remember in this subtle situation is a well-defined concept only in the normal state as $T\to 0$, cannot be studied by ARPES. But the important point is that to the best resolution of ARPES, small Fermi-surface pockets, such as those that would occur if there was an altered translational symmetry and which are suggested by the simple interpretation of the magneto-oscillation experiments, are not observed. This was already known for Bi2212 samples; recently the same conclusion has been arrived at samples of under-doped YBCO \cite{damascelli}. Consistent with the ARPES experiments, specific heat and magnetic susceptibility measurements \cite{loram}, simply interpreted, reveal a density of states near the chemical potential which varies significantly with energy on scales of the order of the Landau level (LL) splittings. The Onsager, Lifshitz-Kosevich theory
assumes that the density of states near the chemical potential is uniform over energies much much larger than the LL splittings.  The deductions from straight-forward application of this theory to deduce the state of the system at $H=0$ from the magnetization or resistivity oscillations at large $H$ are therefore questionable. 

Moreover, diffraction experiment have not observed an altered translational symmetry in Cuprates setting in below $T^*(x)$. A new  symmetry \cite{cmv} has indeed been discovered by polarized neutron diffraction \cite{neutrons} in two different families of cuprates and by dichroic ARPES \cite{kaminski} in another. The phase below $T^*(x)$ breaks Time-reversal symmetry and some reflection symmetries while preserving translational symmetry.

Recently, the ARPES data on  samples of Bi2212 with varying $x$ was organized to suggest that the angle of the Fermi-arc, $\Phi(x,T)$, suitably defined, scales for different $x$ as a function $\Phi(T/T^*(x))$ with $T^*(x)$ \cite{kanigel}, the same temperature below which thermodynamic and transport properties change their characteristic (marginal fermi-liquid) power laws. Importantly for the present discussion, $\Phi(T/T^*(x))$ extrapolates to $0$ as 
$(T/T^*(x)) \to 0$. The extrapolation, if valid, means that the Fermi-surface of the (non-superconducting) pseudogap state is a set of four-points. This interesting possibility suggested earlier \cite{cmv} is far from being definitively established. 
In this note, I discuss how the  observed magneto-oscillations may be reconciled with  a state with fermi-points.  More importantly, I suggest a few definitive experiments to deduce the density of states near the fermi-points and to tell the difference from the small Fermi-surface pockets scenario. 

 A much discussed recent example of a solid with fermi-points (Dirac-points) is Graphene. 
Magneto-oscillations at constant chemical potential, periodic in $1/H$ are observed  and understood in Graphene \cite{zhang}. The fermi-surface deduced is small, given by the ${\bf k}_F$ measured from the Fermi-points.  (Such Dirac points occur also for two-dimensional d-wave superconductors, but the considerations for magneto-oscillation in superconductors have additional subtleties \cite{FT}.) However as explained below, there are important differences between the contours of constant energy near the fermi-points of the Cuprates compared to such contours in Graphene. In the arguments below, I will use methods which reproduce the known results in Graphene (and in conventional metals) and indicate the differences that arise for the case of Fermi-points in underdoped Cuprates.
\\

{\bf Magneto-oscillation Experiments}:

In graphene the dispersion of particles measured from the Dirac-points is $E({\bf k}) = \pm {\bf v_0}\cdot{\bf k}$ leading to a density of states  proportional to the energy $|E|$  measured from the Dirac points. The energy of the Landau levels (LL) is quite different from that in usual metals; it can be calculated directly \cite{guinea} to be 
\bea
\label{LL graphene}
E_p = \pm hv_0|p|^{1/2}(H/2\phi_0)^{1/2}
\eea
$p$ are integers and $\phi_0 = ch/e$ is the quantum of flux. The LL's spread out on either side of the Dirac point as $H$ is increased (with that at $p=0$ tied to zero energy) and their separation is $\propto H^{1/2}$.
Magneto-oscillations at constant chemical potential with oscillations periodic in $1/H$ have been observed, and calculated \cite{sharapov}. At $T=0$, the calculation can be done in a simpler way and for a more general density of states in zero field, so that they are also applicable to the cuprates as well as for experiments at constant particle density.
 
Important features of magneto-oscillations follow simply from the quantization requirements of orbits in a magnetic field and are independent of details of density of states $n(E)$. The physics is especially simple for the case of a two-dimensional solid. Consider that the two-dimensional density of states in $H=0$ varies near the chemical potential $\mu=0$ as
\bea
\label{dos0}
n(E) &=& n_0(E/E_0)^\alpha, ~~|E| \lesssim E_0. 
 \eea
\noindent
For Graphene,  $\alpha =1$. For the pseudogapped cuprates, one model \cite{cmv} gives $\alpha = 1/2$.

Apply now a magnetic field $H$ normal to the solid. Landau levels must form whose degeneracy is given  by the requirement from Bohr-Sommerfeld quantization that the flux enclosed by the semi-classical orbits must equal the quantum of flux.  Since such Landau orbits must fill the two-dimensional space for each Landau level, each Landau level in a sample of unit area normal to the field has a degeneracy $H/\phi_0$. 
 
Let $E_p$ be the energy of the Landau level, $p=\mp1$ referring to the LL just below/just above $E=0$. Then the difference in energy between successive Landau level is given by the requirement that the number of states
\be
 \int_{(E_p-E_{p-1})/2}^{(E_{p+1}-E_p)/2}dE \nu(E) = H/\phi_0 
 \ee. (I have neglected spin splitting of LL's in this discussion.) It follows from Eq.(\ref{dos0}) that the density of states in a field $H$ in the pure limit is
\be
\label{dos}
n_H(E) = \sum_p \frac{H}{\phi_0} \delta(E-E_p).
\ee
with
\be 
\label{landau}
E_p \approx \pm \delta^{-\delta}  |p|^{\delta} E_0 (\frac{1}{n_0E_0})^{\delta}(H/\phi_0)^{\delta}; ~~\delta= (1+\alpha)^{-1}.
\ee
\noindent
One can check that this reproduces the results for the two cases derived more conventionally:(1)The uniform density of states, $\alpha = 0$, where  $\delta=1$, i.e. the Landau levels are equally spaced and $E_p =  p \omega_c$. The case of Graphene \cite{sharapov} where $\alpha = 1$, so that $\delta =1/2$.
 
 Consider, for example, the de Haas-van Alphen effect or oscillatory Magnetization $M(H)$. In two dimensions it is necessary to separately consider the two cases \cite{Miyake}:  (i) The particle density is conserved. The chemical potential $\mu$ lies on a LL and the number of particles in it changes and it moves in energy with field. When it becomes occupied with the maximum number allowed, $\mu$ jumps to the adjacent unfilled  LL, and so on.  (ii) The chemical potential  is fixed and is in general in between the LL's so that all LL's are either fully occupied or completely empty. \\
 
\noindent
{\bf Magneto-oscillations at constant particle density}\\

Consider the case of oscillations at constant particle density, when the chemical potential is always at a LL. For the case of Graphene the chemical potential remains on the $p=0$ LL as the field is varied. Then there can be no oscillations. But as explaned below, no LL may form in the Cuprates at the zero of energy. The oscillations then must take into account the variations in the chemical potential as the magnetic field is varied. Suppose at a field $H$, $p$ LL's are completely filled, each with $H/\phi_0$ particles per unit area. To keep constant density, the LL  closest to $0$ energy is partially filled with density $(H/\phi_0)\nu$ particles per unit area, with $0<\nu<1$. The chemical potential lies in (and moves with) this LL as the total number of LL's changes as $H$ is changed. The density of states per unit-energy and per unit area near . $n(E)$ defined in Eq.(\ref{dos0}) is given in two dimensions by $2\pi dS/dE$, where $S(E)$ is the area in momentum space at constant energy surface $k(E)$.  Then subject to the approximation that $S(E)$ near $E=\mu$ does not change for nearby LL's, $\nu(H)$ is given by the Onsager semi-classical quantization condition.  
\be
\label{quant cond}
2\pi S(\mu) = \frac{H}{\phi_0}  (\nu+p),
\ee
In 2D, Eq. (\ref{quant cond}) is just the condition, in the approximation mentioned, that the density of particles does not change in the magnetic field.  
We can rewrite (\ref{quant cond}), by defining a field $H=H(p)$, where the $p$-th level has $\nu(H)=0^+$, so that
\be
\label{nu1}
\nu(H) = 2\pi S(\mu) \phi_0(1/H-1/H(p)).
\ee

This also satisfies the condition that the $p$-th level has $\nu(H) = 1^-$ at a field $H'(p) = H(p+1) = pH(p)/(p+1)$ where the $p+1$th LL  has $\nu(H)=0^+$. Therefore in ordinary two-dimensional metals with a smooth density of states over several times the LL separation, $\nu$ fluctuates from $0$ to $1$  as a  periodic (triangular) function of $1/H$, with period 

\be
\label{period}
\Omega = \frac{1}{H'(p)}-\frac{1}{H(p)}= \frac{1}{2\pi S(\mu)\phi_0},
\ee
provided $S(\mu)$ is independent of $H$. Note that $2 \pi S(\mu)\phi_0$ is equal to the density of particles in the band integrated from $E=0$ to the energy of the first  LL. 
For an ordinary two-dimensional material, the density of states can be assumed smooth in a region of the several times  the cyclotron energy around the zero-field chemical potential $\mu(0)$ so $S(\mu)$ is very weakly dependent on $H$. The physical properties such as the magnetization then oscillate as $\mu(H)$ does with a period given by $\Omega$.  

Suppose the density of states varies as Eq.(\ref{dos0}) so that the energy of the LL's move away from $\mu(0)$ on both sides as $H$ is increased, as in Eqs. (\ref{dos},\ref{landau}), and that there is no LL at energy $0$.  Then $\mu(H)$ stays in the LL nearest energy $0$  with increasing $H$ while its occupation decreases to $0$, whereupon the total number of LL's decreases but $\mu(H)$ stays again at the the new LL nearest to energy $0$. Its occupation begins to decrease from $1$ as $H$ is further increased, and so on. Physical properties in this case still oscillate because $\nu(H)$ oscillates. For example, 
the magnetization at constant particle density may be calculated, following Ref.(\onlinecite{Miyake}) using
\be
\frac{\partial M}{\partial \rho'}|_\mu = - \frac{\partial \mu}{\partial H}|_{\rho'},
\ee
and integrating with respect to $\rho'$ from the density of the filled LL's to the actual density, or equivalently  from $\nu=0$ to $\nu(H)$.
Magnetization oscillates then with the same period, Eq.(\ref{period}), as $\nu(H)$. 

Another important point is that  the dependence of $S(\mu)$ on $H$ may not be ignored if the density of states at successive LL's in which $\mu(H)$ lies differs significantly.  The dependence of $S$ on $H$ is smooth and does not affect the periodicity condition if $H'(p)-H(p) << H$, the value of the field in which the experiments are done. If $H'(p)-H(p) \lesssim H$, one would be in the condition for quantized Hall effects. In between these two extremes, there should be a slow change of period in $1/H$ with $H$. This is one of the experiments suggested below.\\

\smallskip

{\bf Experiments at constant total particle density but with part of the density in chains and the other in planes}\\

M-O experiments are done so far only in samples in which besides the Cu-O planes, there are also Cu-O chains. The latter do not form Landau-levels but for fixed {\it total} density, as the common chemical potential of the planes and the chains changes, the density in both the planes and the chains changes, with the total held constant. Let $\rho_1$ and $\rho_2$ be the number of particles in the chains and the planes in the area of a unit-cell respectively, with the total density $\rho = \rho_1 + \rho_2$.  Let $n_1(0)$ be the density of states of the chains per unit-energy in an area of the unit-cell near the chemical potential, and $\mu(H)$ be the chemical potential. Then the condition $\partial \rho/\partial H =0$, gives
\be
\label{nu2}
\frac{H}{\phi_0}(\partial \nu/\partial H) +  \frac{\nu}{\phi_0}+ (\frac{p}{\phi_0}+n_1 \partial \mu/\partial H) = 0,
\ee
together with the equation  $\mu(H)= -dE_0(H)/d\rho$, where $E_0(H)$ is the ground state energy in the field $H$. (With $n_1=0$, Eq.(\ref{nu2}) has Eq.(\ref{nu1}) as the solution.) Consider $\mu(H)$ as $H$ is decreased. It is stuck on a given LL and moves with it as its population $\nu(H)$ changes from $0$ to $1$, and this LL is full. Then $\mu(H)$ lies at fixed energy in the continuum of the density of states in the chains till the next LL arrives with its population at $0^+$; the process then repeats. The period of the oscillation of $\nu(H)$ as a function of $1/H$ may be calculated again as in Eq.(\ref{period}) to get
\be
\label{Omega'}
\Omega'  \approx (\delta\rho(E_1)\phi_0)^{-1},
\ee
$(\delta\rho(E_1))$ is the density of particles in the planes {\it and} the chains obtained by integrating the density of states from energy $0$ to the (filled) LL nearest to the energy $0$. This result requires the further approximation that the contribution of the number of particles in the chains varies linearly with $H$, which is consistent with the approximation made earlier to get oscillations periodic in $1/H$ in the case of planes alone.

We may estimate $(\delta\rho(E_1))$ at the typical  field, $50$ Tesla of the experiments. The energy of the first landau level for $E_0 \approx 1000K, N(0) \approx 1 state /eV/unit-cell$ is $E_1\approx 400K$. It has a degeneracy of about $10^{-2}$ states/ unit-cell. In this energy range, the number of states in the chains is $\approx N_1(0)E_1\approx 4\times10^{-2}$states per unit-cell. So $(\delta\rho(E_1)) approx 5\times10^{-2}$states per unit-cell.\\

\smallskip

{\bf Experiments at constant chemical Potential}\\

The experiments in Cuprates are done at constant particle density but for completeness, I give the results for experiments at constant chemical potential using the simple procedure given for instance in Ref. (\onlinecite{Miyake}). For a constant chemical potential, the oscillatory part of the magnetization is also periodic in $1/H$ for any density of states and has the form: 
\be
\label{osc-constMu}
M_{\mu}(H) \propto  {\cal P}((\mu^{1+\alpha}/(A_0E_0)^{1+\alpha}) /(H/\phi_0))
\ee
${\cal P}$ is a triangular periodic function of its argument (in two dimensions). For $\mu \to 0$, the period of oscillations tends to $\infty$. For finite $\mu$ and a constant density of states, the conventional behavior, oscillatory period proportional to $\mu \propto$ the cross-sectional area of the fermi-surface is obtained. For $\alpha =1$, the results are the same as those calculated and for Graphene \cite{sharapov}. In that case $\mu=v_Fk_F$ is to be measured from the Fermi-points. The information from the oscillatory magnetization gives $k_F^2$, the cross-section sectional area at  $\mu$; since this is to be measured from the Fermi-points, it can give a small Fermi-surface indeed. These considerations are in agreement with the magneto-oscillation experiments \cite{zhang, jiang} done in Graphene at constant chemical potential by fixing gate voltage.\\

\smallskip

{\bf The phenomena of Fermi-arcs in Cuprates.}\\
 
\begin{figure}
\centerline{\includegraphics[width=0.6\columnwidth]{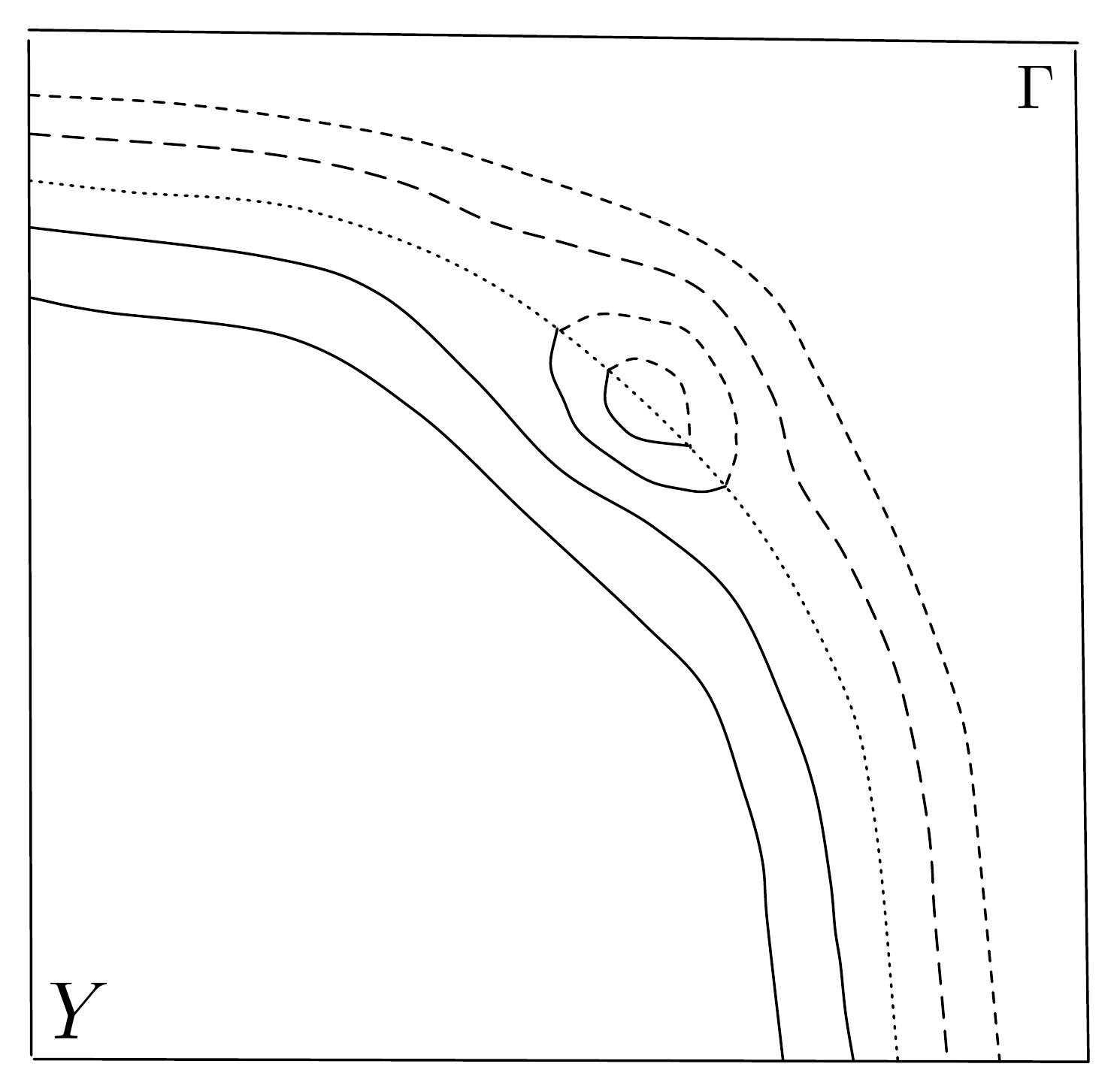}}
\caption{Schematic of a few contours of constant energy which are consistent with the shrinking of "Fermi-arcs" to zero around one of the Fermi-points in the pseduogap phase. The dotted line is the Fermi-surface without a pseudogap; the dashed lines are contours for energies smaller than the chemical potential and the full lines are contours for energies larger than the chemical potential. Note that there is a gap $2D(\theta)$ at the point on the dotted lines where the dashed lines and the full lines touch.}
\label{fig:contours'}
\end{figure}

For Graphene (or d-wave superconductors), there exists points $k=k_F(\theta)$, at $\theta =0$ where $E=\mu=0$. At any finite energy $\pm \Delta E$, there exist closed contours $k_{\Delta E}(\theta)$ around the points at $E=0$. The velocity ${\bf v}({\bf k}) = \nabla_{\bf k}E_{\bf k}$ on any contour is normal to the contour at every point so that in a magnetic field normal to the plane, the Lorentz force drives the electrons along the contour. 
On the other hand, the phenomena of Fermi-arcs, as deduced from ARPES experiments, implies that no closed constant energy contours encircle the Fermi-points for energies below the pseduogap energy at $H=0$. For example, a theory \cite{cmv} giving such arcs has an electronic dispersion
\be
\label{gap}
E({\bf k}) \approx  \epsilon ({\bf k}) -\mu \pm D(\theta, (\epsilon ({\bf k}) -\mu)), ~ for~E({\bf k}) >,< \mu.
\ee
$D(\theta, (\epsilon ({\bf k}) -\mu))\propto \cos^2(2\theta_{\hat{k}})$, so that ${\bf k} =k_F(\theta = 0,\pm \pi/2, \pi)$ are the Fermi-points.  Around such points, contours of constant energy $k_{\Delta E}(\theta)$ are not  closed; there are two arcs of constant energy, one for $\Delta E < \mu$ and one for $\Delta E < \mu$ with a gap $2D(\theta)$.  This is illustrated in Fig.(\ref{fig:contours'}). An important point to note through 
Fig.(\ref{fig:contours'}) is that the change in velocity as energy is increased near the chemical potential is opposite to that for a normal fermi-surface. One would therefore expect the sign of the Hall effect to be opposite the normal sign. This is what is observed \cite{magneto-oscExpts.}.

To get closed orbits in such a situation, magnetic breakdown \cite{blount, shoenberg} is necesary. Magnetic breakdown has always been a misnomer since the effect happens at any field and only the amplitude of the  
oscillations depends on the field. Magnetic breakdown cannot be discussed in semi-classical terms and can be explained in the following way.  Suppose one has a multi-sheeted fermi-surface of a metal with a direct gap $E_0$ between two bands near the fermi-energy  with $E_b=E_0^2/E_f$  smaller or comparable to the cyclotron energy $\omega_c$, where $E_f$ is the fermi-energy . Generally, the ${\bf p}.A({\bf r})$ term in the Hamiltonian has off-diagonal matrix elements between the valence and the conduction bands near the same value of momentum. Here $A({\bf r})$ is the vector potential for the externally applied magnetic field. Then the states in $H$ are linear combination of the states of the two bands and in general form new closed orbits. The amplitude of the oscillations depends on the ratio $E_b/H$ \cite{blount}. The matrix element of coupling between the bands produces a phase-shift $\delta_m$; therefore to complete a semi-classical orbit with a total phase-shift of $2\pi$, the effective area of the orbit is smaller than for closed orbits. Since (in the leading approximation) $\delta_m \propto H$ as is the phase-shift moving along the arcs of constant energy, no change in the dependence of the oscillatory phenomena with $H$ is entailed. The degeneracy of the effective Landau orbitals is now $H/\phi_m$ with $\phi_m=\phi_0(1-\delta_m/2\pi)$. Calculating $\delta_m$ requires detailed knowledge of the unperturbed band wave-functions and is complicated. 

Consider now the situation of the electronic structure of 
Cuprates with the four fermi-points as described above. The obvious "magnetic breakdown"  is just the transition across the gap 
$2D({\bf \hat{k}})$.  An important point to note is that since the closed orbits from states in the pseudogap energy region are formed due to the perturbative admixture of states above and below the chemical potential, no $p=0$ LL at the chemical potential can exist. This is an important distinction from the LL's in Graphene.\\

\smallskip

{\bf Estimation of the Period of Oscillations}\\

It is important to show that the expected period from the above analysis is similar to the experiments. The period of the oscillations in the experiments \cite{jaudet} done in a field around 50 Tesla in $YBa_2CU_3O_{6.5}$ is $\Omega \approx 2\times 10^{-3} Tesla^{-1}$.  Let us compare the observed $\Omega$ with that one may  estimate using Eq. (\ref{Omega'}).  Taking $N(0) \approx 1/(ev-unit-cell)$, where the unit-cell has an area $A_0 \approx 16 Angstroms^2$, and $N(0)$ includes the contribution of both Cu-O planes in the unit-cell and including the effect of the chains as estimated following Eq.\ref{Omega'}, the period calculated is $\approx 3\times 10^{-3} Tesla^{-1}$. This is necessarily a crude estimate. Beside the uncertainty of the parameters to factors of $O(2)$, no account of the phase-shift due to magnetic breakdown necessary to form closed orbits has been taken into account.  One may however conclude that the  estimate does not rule out the consistency of the magneto-oscillation results with the "Fermi-arcs" turning to Fermi-points as $T\to 0$ suggested by other experiments. Note the important fact that the sign of the Hall effect by these consideration is opposite to that one would get from the conventional interpretation with small Fermi-surface pockets brought about by a new translational symmetry. This has been a principal mystery about the experimental results.

Yet another point to note is that although the LL separation is an order of magnitude larger than normal, the spin-splitting is the normal value. In such a case it may be hard to resolve the spin-splitting except if the 
$Q$ of the oscillations is very high. Experiments rotating the direction of the field to discern the spin-splitting have failed  to see any \cite{harrison}.

{\bf  Experiments to 
Distinguish between Fermi-Points and Conventional Small Fermi-surface}\\

To distinguish the ideas of this paper from the conventional small Fermi-pockets scenario, one needs to have experiments, which will yield {\it qualitatively} different results for the two cases. Three experiments are proposed here.

(1) The most direct and the most important experiment to do is infra-red absorption as a function of magnetic field.
For the conventional situation the successive absorption peaks due to transitions between the unfilled LL's  to the un-filled LL's  are separated by the cyclotron energy $\propto H$.  For Fermi-points, as derived in Eq. ({\ref{landau}), they are separated by $ H^{\delta}$. For Graphene, this behavior with $\delta=1/2$ has already been observed \cite{jiang}. For Cuprates, the dispersion of Eq.(\ref{gap}) gives  $\delta = 2/3$ is expected. In general any $\delta<1$ can only be consistent with a density of states which goes to zero at $\mu(0)$, (apart from the effects due to impurities). 

(2) In the underdoped cuprates, the oscillatory period should $\to \infty$ if the experiment can be done such that at zero field the density of states at the chemical potential is $0$, as suggested by the shrinking "fermi-arcs". This requires doing magneto-oscillation experiments in cuprates at constant chemical potential. There should also be a quantitative difference in experiments on cuprates without chains.

(3) The slow variation of the oscillation frequency due the field that has already been discussed. The magnitude of the effect is that a factor of 2 change in the field in the experiments should increase the period by about
$10\%$.\\

{\bf Concluding Remarks}\\

A valid theory of Cuprates must be based on a single set of ideas from which all the universal observed properties should be derivable. The predicted quantum-criticality in Cuprates which may now be regarded as definitively established experimentally is the central aspect of such a set of ideas. From the quantum-critical point in the phase diagram  emanate a broken time-reversal symmetry state with peculiar properties, a quantum-critical region with scale-invariant universal transport and thermodynamic properties truncated at low temperature by a superconducting phase, and a Fermi-liquid region. It is therefore important that the observed magneto-oscillations in the broken symmetry state be well understood within the same framework. In this paper, I have tried to show that if the broken symmetry state (the pseudogap state) has a "normal" ground state with four fermi-points, the observed magneto-oscillations can follow. Unlike the normal state of graphene, the four-fermi-point ground state of Cuprates requires a magnetic-field induced formation of closed orbits to have magneto-oscillations from the two-dimensional states and the inclusion of the particles in the chains to get reasonable quantitative agreement with the observed period of oscillations. Since neither the occurrence of the four-fermi point ground state nor the fulfilling of these requirements is 
proven experimentally beyond doubt, it is important to have further experiments. Toward resolving the ground state electronic structure  in the pseudogap phase of the Cuprate, I have proposed a few new experiments in a magnetic field. Of these, infra-red absorption in a magnetic field would be the most revealing of a ground state with four fermi-points.

{\bf Acknowledgements} I have had useful discussions on the subject matter of this paper with Elihu Abrahams, Vivek Aji, Arcadi Shehter, and Suchitra Sebastian. I wish to thank Elihu Abrahams for detailed comments on the manuscript.

\end{document}